\documentstyle[aasms4,12pt]{article}
\begin{document}

\title{Evolution of O Abundance Relative to Fe}
\author{Y.-Z. Qian\altaffilmark{1} and G. J. Wasserburg\altaffilmark{2}}
\altaffiltext{1}{School of Physics and Astronomy, University of
Minnesota, Minneapolis, MN 55455; qian@physics.umn.edu.}
\altaffiltext{2}{The Lunatic Asylum,
Division of Geological and Planetary Sciences, California
Institute of Technology, Pasadena, CA 91125.}
\centerline{\today}
\pagestyle{myheadings}
\markright{(O/Fe)-\today}

\begin{abstract}
We present a three-component mixing model for the evolution of O abundance 
relative to Fe, taking into account the contributions 
of the first very massive stars (with masses of $\gtrsim 100\,M_\odot$) 
formed from Big Bang 
debris. We show that the observations of O and Fe abundances in metal-poor
stars in the Galaxy by Israelian et al. (1998) and Boesgaard et al. (1999)
can be well represented both qualitatively and quantitatively by this model.
We use the representation of the number ratios (O/Fe) versus 1/(Fe/H).
In this representation, if there is only a single source with a fixed 
production ratio of O to Fe beyond a certain point, the subsequent evolution 
of (O/Fe) is along a straight line segment.
Under the assumption of an initial Fe ([Fe/H]~$\sim -3$) and O inventory 
due to the prompt production by the first very massive stars, the data of
Israelian et al. (1998) and Boesgaard et al. (1999) at
$-3\lesssim$~[Fe/H]~$\lesssim -1$ are interpreted to result from the addition
of O and Fe only from type II supernovae (SNII) to the prompt inventory.
At [Fe/H]~$\gtrsim -1$, 
SNII still contribute O while both SNII and type Ia
supernovae contribute Fe. During this later stage, (O/Fe) sharply
drops off to an asymptotic value of $\sim 0.8$~(O/Fe)$_{\odot}$.
The value of (O/Fe) for the prompt inventory at [Fe/H]~$\sim -3$
is found to 
be (O/Fe)~$\sim 20$~(O/Fe)$_\odot$.  This result suggests that 
protogalaxies with low ``metallicities'' should exhibit high values of
(O/Fe). The C/O ratio produced by the first very massive stars is expected 
to be $\ll 1$ so that all the C should be tied up as 
CO and that C dust and hydrocarbon compounds should be 
quite rare at epochs corresponding to [Fe/H]~$\lesssim -3$.
\end{abstract}
\keywords{Galaxy: abundances --- Galaxy: evolution --- supernovae: general}

\section{Introduction}
In a recent theoretical study of Fe and $r$-process abundances
in the early Galaxy by Wasserburg \& Qian (2000a), it was
proposed that the first stars formed after the Big Bang were very
massive ($\gtrsim 100\,M_\odot$) and promptly enriched the interstellar
medium (ISM) to [Fe/H]~$\sim -3$, at which metallicity formation of regular
stars (with masses of $\sim 1$--$60\,M_\odot$) took over. Subsequent Fe
enrichment was provided initially by a subset of Type II supernovae (SNII).
The interpretations of Wasserburg \& Qian (2000a) were based on
observations of metal-poor stars in the Galaxy by Gratton \& Sneden (1994),
McWilliam et al. (1995), McWilliam (1998), and Sneden et al. (1996, 1998).
The critical metallicity [Fe/H]~$\sim -3$ deduced by Wasserburg \& Qian
(2000a) for transition from formation of very massive stars to regular stars
is in remarkable coincidence with the lower bound of [Fe/H]~$\sim -2.7$
observed in damped Ly$\alpha$ systems over a wide range in redshift
(e.g., Prochaska \& Wolfe 2000).
Furthermore, Wasserburg \& Qian (2000b) showed that the
[Fe/H] of damped Ly$\alpha$ systems can be explained in a straightforward
manner by addition of Fe from SNII to
the prompt inventory provided by the first very massive stars 
and that the evolution of [Fe/H] with
redshift reflects the chronology of turn-on of protogalaxies.

The first very massive stars would certainly produce elements other than Fe.
One of the issues to be resolved is the general problem of the
nucleosynthetic signatures of these stars.
This problem was discussed in the early work by Ezer \& Cameron (1971).
More recently, studies on condensation
of Big Bang debris (e.g., Bromm, Coppi, \& Larson 1999) indicated that
protostellar aggregates
of $\sim 100\,M_\odot$ may be formed.
Evolution of zero-metallicity stars of $\gtrsim 100\,M_\odot$ was
discussed by Fryer, Woosley, \& Heger (2000). Calculations by Heger,
Woosley, \& Waters (2000) suggest that C, N, O, Mg, Si, and Fe are produced
in large amounts by these stars. The question is how the theoretical
models of nucleosynthesis in the first very massive stars formed from
Big Bang debris are related to the observations.
Another issue concerns the Fe
enrichment by Type Ia supernovae (SNIa). It was assumed by Wasserburg \&
Qian (2000a,b) that SNIa started to contribute Fe at [Fe/H]~$\sim -1$ and
were responsible for $\sim 2/3$ of the solar Fe inventory. In a
self-consistent
Fe enrichment model, the history of SNIa Fe contribution needs
to be examined more carefully based on observations.

In this paper we present a phenomenological model for the
evolution of O abundance relative to Fe. Our approach differs from previous
studies (e.g., Timmes, Woosley, \& Weaver 1995) in the following respects.
Previous studies rely on many inputs such as the nucleosynthetic yields
predicted by various models of SNII and SNIa and some description of star
formation. In the approach here we attempt to identify a chemical
evolution model that can account for the observed evolution of O abundance
relative to Fe in terms of several key parameters. We hope that this 
approach will give a simple yet complete interpretation of the data.
The usual representation of the O and Fe data is to display [O/Fe]
against [Fe/H]. In the approach here we use (O/Fe) versus 1/(Fe/H), with
the round brackets indicating the number ratios.
If there is only a single source with a fixed production ratio of O to Fe
beyond a certain point, the subsequent evolution of (O/Fe) as a function
of 1/(Fe/H) is along a straight line segment. As will be shown,
this representation also facilitates a simple geometric means to describe
the evolution of O and Fe abundances in the general case.
As the first very massive stars are thought to be a
source of O and Fe (Heger et al. 2000), the value of (O/Fe) at
[Fe/H]~$\lesssim -3$ should serve
as a nucleosynthetic signature of these stars, thus addressing the first
issue raised above. Subsequent to the prompt enrichment 
by the very massive stars,
SNII are essentially the only source of O, while Fe enrichment was 
provided by SNII initially and 
augmented by the contribution from SNIa only at a later time.
Thus the evolution of O abundance relative to Fe at [Fe/H]~$>-3$
provides a means of addressing questions concerning the onset and
nature of SNIa Fe contribution. 

Based on the observations of
metal-poor stars in the Galaxy by
Israelian, Garcia Lopez, \& Rebolo (1998) and Boesgaard et al. (1999),
the value of (O/Fe) resulting from the prompt production by
the first very massive stars is $\sim 20$~(O/Fe)$_\odot$.
Comparison of our model with the same observations
gives a history of SNIa Fe contribution that is
consistent with the results of previous studies
(e.g., Timmes et al. 1995) 
and with the assumptions of Wasserburg \& Qian
(2000a,b). The average production of O relative to Fe for SNII is
found to be (O/Fe)$_{\rm II}\sim 3$~(O/Fe)$_\odot$, which is $\sim 7$
times lower than that for the very massive stars. Our model is
presented in \S2 and compared with observations in \S3. Discussion and
conclusions are given in \S4.

\section{Model for Evolution of O Abundance Relative to Fe}
For simplicity, we focus on the evolution of O and Fe abundances in a
closed system with a fixed mass. This mass can be stored in gas and stars.
The initial state P at time $t=0$ contains gas only. At $t>0$, the total
mass of gas $M_{\rm gas}(t)$ and that of stars $M_{\rm star}(t)$ satisfy
\begin{equation}
M_{\rm gas}(t)+M_{\rm star}(t)=M_{\rm gas}(0).
\label{mass}
\end{equation}
We ignore the return of O and Fe to the gas by stars that 
are not related to SNII or SNIa and
consider that stars sample the composition of the gas at the time of
their birth. Thus the mass fraction $X_{\rm E}(t)$
of the element E in the gas is governed by
\begin{equation}
X_{\rm E}(t)M_{\rm gas}(t)+
\int_0^t X_{\rm E}(t'){dM_{\rm star}(t')\over d\,t'}d\,t'=
\int_0^t[P_{\rm E,II}(t')+P_{\rm E,I}(t')]d\,t',
\label{pxe}
\end{equation}
where $P_{\rm E,II}(t)$ and $P_{\rm E,I}(t)$ are 
the total rates for production of the element by SNII and SNIa.
 
Differentiating equation (\ref{pxe}) with respect to $t$ gives 
\begin{equation}
{d\,X_{\rm E}\over d\,t}M_{\rm gas}(t)=
P_{\rm E,II}(t)+P_{\rm E,I}(t),
\label{dxe}
\end{equation}
where equation (\ref{mass}) has been used.
Equation (\ref{dxe}) can be integrated to give
\begin{equation}
X_{\rm E}(t)=X_{\rm E}(0)+
\int_0^t{P_{\rm E,II}(t')\over M_{\rm gas}(t')}d\,t'
+\int_0^t{P_{\rm E,I}(t')\over M_{\rm gas}(t')}d\,t'.
\label{xe}
\end{equation}
Note that the integrands in equation (\ref{xe}) only depend on
the production rates per unit mass of gas.
As the mass fraction of H in the gas is nearly constant during its
evolution, $X_{\rm E}(t)$ is proportional to the number ratio 
(E/H). So equation (\ref{xe}) can be rewritten as
\begin{equation}
{\rm (E/H)}={\rm (E/H)}_{\rm P}+{\rm (E/H)}_{\rm II}+{\rm (E/H)}_{\rm I},
\end{equation}
where (E/H)$_{\rm P}$ is the inventory for the initial state P, 
and (E/H)$_{\rm II}$ 
and (E/H)$_{\rm I}$ represent the integral of the production rate
per unit mass of gas over time for SNII and SNIa.

We consider that
the first very massive stars (with masses of $\gtrsim 100 M_\odot$) 
formed from Big Bang debris
provided an inventory of Fe ([Fe/H]~$\sim -3$) and O.
Frequently we use ``prompt inventory'' or ``prompt production''
when referring to this inventory of Fe and associated elements.
It is considered that only very massive stars could form until this 
level of Fe and associated elements was reached. The onset
of formation of regular stars (with masses of $\sim 1$--$60\,M_\odot$)
corresponds to [Fe/H]~$\sim -3$.
This defines the initial state P of the system.
Subsequently, SNII that are related to the formation of 
regular massive stars (with masses of
$\sim 10$--$60\,M_\odot$) began to contribute O and
Fe. At a still later time (see \S3),
SNIa also began to contribute additional
Fe. It is considered (e.g., Timmes et al. 1995)
that SNIa essentially produce no O. So we assume a production ratio
(O/Fe)$_{\rm I}=0$ for SNIa. The O and
Fe yields may vary significantly between individual SNII (see \S4). 
However, the O or Fe contribution from all SNII may be
lumped together so long as we are concerned with timescales of, e.g.,
$\gtrsim 10^8$~yr (see \S4). Furthermore, we will 
assume a constant average production ratio (O/Fe)$_{\rm II}$
for SNII.

The value of (O/Fe) for the initial state P resulting from the prompt
production by the first very massive stars is taken to be a constant 
(O/Fe)$_{\rm P}$. Thus (O/Fe)$_{\rm P}$ represents the production
ratio for the prompt source, (O/Fe)$_{\rm II}$ for SNII, and
(O/Fe)$_{\rm I}=0$ for SNIa. The whole evolution of O and Fe abundances
can then be described by a three-component mixing model.
In consideration of the time sequence indicated above, there is a domain
where the model reduces to  only two components,
the prompt source
and SNII (no SNIa). In this domain,
the number ratio of O to Fe in the gas is
\begin{equation}
{\rm (O/Fe)}={{\rm (O/H)}_{\rm P}+{\rm (O/H)}_{\rm II}\over
{\rm (Fe/H)}_{\rm P}+{\rm (Fe/H)}_{\rm II}}
={{\rm (O/Fe)}_{\rm P}{\rm (Fe/H)}_{\rm P}+
{\rm (O/Fe)}_{\rm II}{\rm (Fe/H)}_{\rm II}\over {\rm (Fe/H)}},
\label{ofe}
\end{equation}
where ${\rm (Fe/H)}={\rm (Fe/H)}_{\rm P}+{\rm (Fe/H)}_{\rm II}$.
Equation (\ref{ofe}) can be rewritten as
\begin{equation}
{\rm (O/Fe)}={\rm (O/Fe)}_{\rm II}+
[{\rm (O/Fe)}_{\rm P}-{\rm (O/Fe)}_{\rm II}]
{{\rm (Fe/H)}_{\rm P}\over{\rm (Fe/H)}}.
\label{ofe2}
\end{equation}
Thus the ratio (O/Fe) is
a linear function of (Fe/H)$_{\rm P}$/(Fe/H) with a slope of
[(O/Fe)$_{\rm P}-{\rm (O/Fe)}_{\rm II}$].
The evolution of (O/Fe) in this regime would be along the line segment PII
in Figure 1. Point P is at (O/Fe)~=~(O/Fe)$_{\rm P}$ and
(Fe/H)$_{\rm P}$/(Fe/H)~=~1, and represents the initial state P.
Point II is at (O/Fe)~=~(O/Fe)$_{\rm II}$
and (Fe/H)$_{\rm P}$/(Fe/H)~=~0, and represents the result after an infinite
number of SNII (eq. [\ref{ofe2}]).
If SNII cease to occur after contributing finite amounts of Fe and O,
then the end point for this part of the evolution would lie somewhere on
the line segment PII.

If the SNII contributions were to cease and SNIa were to begin to
contribute Fe at the state corresponding to point A on
the line segment PII in Figure 1,
then the evolution of (O/Fe)
would again reduce to a two-component mixing
model with one member being state A and the
other being SNIa. In this simplified case (no
concurrent production by SNII), the evolution of (O/Fe) beyond point A
would be along the line segment AI in Figure 1 corresponding to
\begin{equation}
{\rm (O/Fe)}
={\rm (O/Fe)}_{\rm A}{{\rm (Fe/H)}_{\rm A}\over{\rm (Fe/H)}},
\label{ofe1}
\end{equation}
where ${\rm (Fe/H)}={\rm (Fe/H)}_{\rm A}+{\rm (Fe/H)}_{\rm I}$,
and the subscript ``A'' stands for quantities at point A.
Point I in Figure 1 is at (O/Fe)~=~(O/Fe)$_{\rm I}=0$ and
(Fe/H)$_{\rm P}$/(Fe/H)~=~0, and
represents the result after an infinite number of SNIa (eq. [\ref{ofe1}]).

As SNII still occur after SNIa start to contribute Fe at point A,
the actual evolution of (O/Fe) follows a curve between the
line segments AII and AI in Figure 1. This is a
three-component mixing regime involving state A,
SNII, and SNIa. An important parameter for this regime is the rate 
per unit mass of gas for
Fe production by SNIa relative to SNII:
\begin{equation}
\gamma\equiv{d\,{\rm (Fe/H)}_{\rm I}/d\,t\over 
d\,{\rm (Fe/H)}_{\rm II}/d\,t}.
\end{equation}
For a constant $\gamma$, we have
\begin{equation}
{\rm (O/Fe)}={{\rm (O/Fe)}_{\rm II}\over\gamma+1}+
\left[{\rm (O/Fe)}_{\rm A}-{{\rm (O/Fe)}_{\rm II}\over\gamma+1}\right]
{{\rm (Fe/H)}_{\rm A}\over{\rm (Fe/H)}},
\label{ofec}
\end{equation}
corresponding to the line segment AC in Figure 1a. Point C is at
(O/Fe)~=~(O/Fe)$_{\rm II}/(\gamma+1)$ and
(Fe/H)$_{\rm P}$/(Fe/H)~=~0, and represents
the result after an infinite number of SNII and SNIa (eq. [\ref{ofec}]).
It can be seen that the line segments AII and AI correspond to the limits
$\gamma=0$ and $\infty$.

In general, $\gamma$ is a function of time $t$. We can discuss
the evolution of (O/Fe) as a function of (Fe/H)$_{\rm P}$/(Fe/H)
for the case of a time-dependent $\gamma$
by calculating $d\,{\rm (O/Fe)}/d\,t$ and
$d\,[{\rm (Fe/H)}_{\rm P}/{\rm (Fe/H)}]/d\,t$. From the rates for addition to
the total O and Fe inventory in the gas,
\begin{equation}
{d\,{\rm (O/H)}\over d\,t}={d\,{\rm (O/H)}_{\rm II}\over d\,t}=
{\rm (O/Fe)}_{\rm II}{d\,{\rm (Fe/H)}_{\rm II}\over d\,t},
\label{odot}
\end{equation}
and
\begin{equation}
{d\,{\rm (Fe/H)}\over d\,t}=
{d\,{\rm (Fe/H)}_{\rm II}\over d\,t}+
{d\,{\rm (Fe/H)}_{\rm I}\over d\,t}=
[\gamma(t)+1]{d\,{\rm (Fe/H)}_{\rm II}\over d\,t},
\label{fedot}
\end{equation}
we obtain
\begin{equation}
S\equiv {d\,{\rm (O/Fe)}\over d\,[{\rm (Fe/H)}_{\rm P}/{\rm (Fe/H)}]}=
{{\rm (O/Fe)}-\{{\rm (O/Fe)}_{\rm II}/[\gamma(t)+1]\}\over
{\rm (Fe/H)}_{\rm P}/{\rm (Fe/H)}}.
\label{s}
\end{equation}
Figure 1b gives a simple geometric interpretation of equation (\ref{s}).
At an arbitrary point B beyond point A, the tangent $S$ of the curve for
(O/Fe) is the same as the slope of the line segment BC$'$.
Point C$'$ is at (O/Fe)~=~(O/Fe)$_{\rm II}/(\gamma_{\rm B}+1)$
and (Fe/H)$_{\rm P}$/(Fe/H)~=~0, and represents
the result after an infinite number of SNII and SNIa for a constant
$\gamma$ equal to the value $\gamma_{\rm B}$ at point B (eq. [\ref{ofec}]).

Equation (\ref{s}) also gives
\begin{equation}
{d\,S\over d\,[{\rm (Fe/H)}_{\rm P}/{\rm (Fe/H)}]}=
{1\over{\rm (Fe/H)}_{\rm P}/{\rm (Fe/H)}}
{{\rm (O/Fe)}_{\rm II}\over [\gamma(t)+1]^2}
{d\,\gamma\over d\,[{\rm (Fe/H)}_{\rm P}/{\rm (Fe/H)}]}.
\label{ds}
\end{equation}
According to equation (\ref{ds}), if $\gamma(t)$ increases with time, i.e.,
$d\,\gamma/d\,[{\rm (Fe/H)}_{\rm P}/{\rm (Fe/H)}]<0$, then
$d\,S/d\,[{\rm (Fe/H)}_{\rm P}/{\rm (Fe/H)}]<0$
and the corresponding 
curve for (O/Fe) is concave downward. On the other hand,
if $\gamma(t)$ decreases with time, i.e.,
$d\,\gamma/d\,[{\rm (Fe/H)}_{\rm P}/{\rm (Fe/H)}]>0$, then
$d\,S/d\,[{\rm (Fe/H)}_{\rm P}/{\rm (Fe/H)}]>0$
and the corresponding
curve for (O/Fe) is concave upward. Figure 1c illustrates the
trend of (O/Fe) for some hypothetical
time evolution of $\gamma(t)$
specified by $d\,\gamma/d\,t\equiv\dot{\gamma}$.
In the simple case where $\gamma(t)$ monotonically increases
with time, the curve for (O/Fe) drops more and
more steeply in approaching (Fe/H)$_{\rm P}$/(Fe/H)~=~0 (Fig. 1b).
In any case,
the value of (O/Fe) corresponding to (Fe/H)$_{\rm P}$/(Fe/H)~=~0
is (O/Fe)$_{\rm II}/(\gamma_\infty+1)$, with
$\gamma_\infty$ being the asymptotic value of $\gamma(t)$ after an infinite
amount of time. Note that in the above treatment, the evolution of
(O/Fe) is directly related to the Fe inventory (Fe/H). The time $t$ only
serves as an implicit parameter that has the same sense of increase
as (Fe/H). 

\section{Comparison with Observations}
We now proceed to compare the phenomenological model presented in \S2 with 
the observational data on the evolution of O abundance relative to Fe.
We take [Fe/H]$_{\rm P}=\log{\rm (Fe/H)}_{\rm P}-\log{\rm (Fe/H)}_\odot=-3$
to correspond to the initial state P resulting from the prompt production 
by the first very massive stars. Here the subscript ``$\odot$'' indicates 
quantities pertaining to the sun.
The data from the observations of metal-poor stars (mostly unevolved halo
stars) in the Galaxy by Israelian et al. (1998) and 
Boesgaard et al. (1999) are
shown as (O/Fe)/(O/Fe)$_\odot=10^{\rm [O/Fe]}$ versus
(Fe/H)$_{\rm P}$/(Fe/H)~$=10^{{\rm [Fe/H]}_{\rm p}-{\rm [Fe/H]}}$
in Figure 2a. The data at (Fe/H)$_{\rm P}$/(Fe/H)~$<0.066$ are shown 
without error bars in the inset.
Note that the data of Boesgaard et al. (1999)
have been modified to match the scale of stellar parameters used by
Israelian et al. (1998) (see Israelian, Garcia Lopez, \& Rebolo 2000).
Figure 2a shows that the evolution of (O/Fe) 
follows a linear trend between the initial
state P and an intermediate state at (Fe/H)$_{\rm P}$/(Fe/H)~$\sim 0.02$.
The evolution beyond the intermediate state follows a much steeper curve.
These evolution trends are expected from the model, with the intermediate
state corresponding to state A at the onset of SNIa Fe contribution.

For a quantitative comparison of the model trajectory with the data
on the evolution of O abundance relative to Fe, 
we assume a time parameterization where
the rate per unit mass of gas for Fe production by SNII 
is constant over Galactic history. This rate can be
written as
\begin{equation}
{d\,{\rm (Fe/H)}_{\rm II}\over d\,t}=
{\alpha {\rm (Fe/H)}_\odot\over t_{\rm SSF}},
\label{rfe2}
\end{equation}
where $\alpha$ is the fraction of the solar Fe inventory contributed by 
SNII and $t_{\rm SSF}$ is the time of solar system formation (with $t=0$
corresponding to the initial state P). The occurrence of SNIa is 
considered to depend on the accretion of matter onto a white dwarf
from its binary companion. The formation of white dwarfs requires 
long-term stellar evolution and thus SNIa cannot occur early in Galactic 
history. We assume that SNIa Fe production begins 
to occur at state A and then increases to a constant value relative to SNII 
Fe production over a timescale $\hat\tau$ (it will 
be shown that the results are not sensitive to the choice of $\hat\tau$).
This is represented by the following expression of the rate per unit mass
of gas for Fe production by SNIa relative to SNII:
\begin{equation}
\gamma(t)=\gamma_\infty
\left[1-\exp\left(-{t-t_{\rm A}\over\hat\tau}\right)\right]
\theta(t-t_{\rm A}),
\label{rfe1}
\end{equation}
where $t_{\rm A}$ is the time corresponding to state A, and
$\theta(t-t_{\rm A})$ is a step function with
its value being unity at $t\geq t_{\rm A}$ and zero otherwise.
Under the above assumptions, the O and Fe inventory in the gas is
\begin{equation}
{\rm (O/H)}={\rm (O/Fe)}_{\rm P}{\rm (Fe/H)}_{\rm P}
+\alpha {\rm (O/Fe)}_{\rm II}{\rm (Fe/H)}_\odot(t/t_{\rm SSF}),
\label{oh}
\end{equation}
and
\begin{eqnarray}
{\rm (Fe/H)}&=&{\rm (Fe/H)}_{\rm P}+
\alpha {\rm (Fe/H)}_\odot(t/t_{\rm SSF})\nonumber\\
&+&\alpha\gamma_\infty {\rm (Fe/H)}_\odot\left\{
{t-t_{\rm A}\over t_{\rm SSF}}-{\hat\tau\over t_{\rm SSF}}
\left[1-\exp\left(-{t-t_{\rm A}\over\hat\tau}\right)\right]\right\}
\theta(t-t_{\rm A}).
\label{feh}
\end{eqnarray}

After choosing a set of parameters (O/Fe)$_{\rm P}$, (Fe/H)$_{\rm P}$,
(Fe/H)$_{\rm A}$, $\alpha$, and $\hat\tau/t_{\rm SSF}$,
we can obtain (O/Fe) as a function of
(Fe/H)$_{\rm P}$/(Fe/H) from equations (\ref{oh}) and (\ref{feh}).
The quantities ${\rm (O/Fe)}_{\rm II}$, $t_{\rm A}$, and
$\gamma_\infty$ in these equations are given by
\begin{equation}
{\rm (O/Fe)}_{\rm II}={1\over\alpha}\left[{\rm (O/Fe)}_\odot-
{\rm (O/Fe)}_{\rm P}{{\rm (Fe/H)}_{\rm P}\over{\rm (Fe/H)}_\odot}\right],
\label{ofeii}
\end{equation}
\begin{equation}
{t_{\rm A}\over t_{\rm SSF}}=
{{\rm (Fe/H)}_{\rm A}-{\rm (Fe/H)}_{\rm P}\over\alpha {\rm (Fe/H)}_\odot},
\label{ta}
\end{equation}
and
\begin{equation}
\alpha\gamma_\infty=
{1-\alpha-[{\rm (Fe/H)}_{\rm P}/{\rm (Fe/H)}_\odot]\over
1-(t_{\rm A}/t_{\rm SSF})-(\hat\tau/t_{\rm SSF})
\{1-\exp[-(t_{\rm SSF}-t_{\rm A})/\hat\tau]\}}.
\label{fehs}
\end{equation}
Note again that the time $t$ (measured in units of $t_{\rm SSF}$)
is introduced as
an implicit parameter to be eliminated from the final results.
The form of this parameterization does not significantly 
affect these results.

As a numerical example, we take (O/Fe)$_{\rm P}=21$~(O/Fe)$_\odot$,
(Fe/H)$_{\rm P}=10^{-3}$~(Fe/H)$_\odot$,
(Fe/H)$_{\rm A}=0.05$~(Fe/H)$_\odot$ 
[corresponding to (Fe/H)$_{\rm P}$/(Fe/H)$_{\rm A}=0.02$],
$\alpha=1/3$ for the fraction of solar Fe inventory contributed by SNII, 
and $\hat\tau/t_{\rm SSF}=0.1$.
The corresponding trajectory of
(O/Fe) as a function of (Fe/H)$_{\rm P}$/(Fe/H)
is shown as the solid curve in Figures 2a, 2b, and 2c. These figures
cover a series of nested intervals in (Fe/H)$_{\rm P}$/(Fe/H) to expose
the complete range of results. For reference, 
the same results (solid curve)
are also shown in the conventional representation of [O/Fe] versus 
[Fe/H] in Figure 2d. The data of Edvardsson et al. (1993) 
for Galactic disk stars 
are included in Figures 2c and 2d to show the trend near and beyond
the solar point (circle). It is evident that
the model provides both a good qualitative and
quantitative description of all the data. 
There is some scatter of the data
relative to the solid curve from the model over almost the entire range of 
(Fe/H)$_{\rm P}$/(Fe/H), especially in the range
$0.03\lesssim$~(Fe/H)$_{\rm P}$/(Fe/H)~$\lesssim 0.1$. However, we do not
consider this to be critical. 

To test the sensitivity of the results to the choice of parameters,
we have varied $\hat\tau/t_{\rm SSF}$, (Fe/H)$_{\rm A}$, or
$\alpha$ one at a time while 
keeping all the other parameters the same as in the above example.
The long-dashed curve in Figure 2c is obtained
by taking $\hat\tau/t_{\rm SSF}=1$ and
the short-dashed curve by taking (Fe/H)$_{\rm A}=0.1$~(Fe/H)$_\odot$ 
[corresponding to (Fe/H)$_{\rm P}$/(Fe/H)$_{\rm A}=0.01$]. 
These two variations seem to
have little effect on the description of the data. The only significant 
change is that the value of (O/Fe)/(O/Fe)$_\odot$ in the infinite future 
decreases from 0.8 for the solid curve to 0.36 for the long-dashed curve
due to the tenfold increase in $\hat\tau/t_{\rm SSF}$.
Changing $\alpha$ from 1/3 to 1/4 only slightly improves the agreement
with the data in the range 
$0.03\lesssim$~(Fe/H)$_{\rm P}$/(Fe/H)~$\lesssim 0.1$ (not shown). 
It would be difficult to obtain a fit to the data for $\alpha$ significantly 
greater than 1/3.

The value of (O/Fe) for the initial state P appears to be 
(O/Fe)$_{\rm P}\sim 20$~(O/Fe)$_\odot$ from the existing data.
This clearly shows that the contributors to the prompt Fe
inventory have also made major contribution to O and very
possibly to C, Mg, and Si. The value (O/Fe)$_{\rm P}$ is considered here
to be the result from the production by the first very massive stars 
(with masses of $\gtrsim 100\,M_\odot$) 
and not by SNII (see \S4). The results from complete
models of nucleosynthesis in these stars should provide a guide 
as to what may be expected for the other elements. 
The average production of O relative to Fe for SNII is
(O/Fe)$_{\rm II}\sim 3$~(O/Fe)$_\odot$, which is $\sim 7$
times lower than (O/Fe)$_{\rm P}$ for the very massive stars.
The high value of
(O/Fe)$_{\rm P}$ required by the data 
implies that even after the major
addition of Fe and O from SNII, the total O inventory at the onset
of SNIa Fe contribution may still have a significant component 
from the prompt source.
For example, at (Fe/H)$_{\rm P}$/(Fe/H)~$\sim 0.02$ 
([Fe/H]~$\sim -1.3$),
$\sim 10$\% of the O inventory but only $\sim 2$\% of 
the Fe inventory are from the prompt source.

We are further concerned with the possibility that prior to the formation
of very massive stars with masses of $\gtrsim 100\,M_\odot$, 
there may have been 
a separate generation of supermassive stars with masses of 
$\gtrsim 10^3$--$10^4\,M_\odot$.
Such objects may produce peculiar values of (O/Fe)
in the data around (Fe/H)$_{\rm P}$/(Fe/H)~$\sim 1$.
We have sought to find an indication of this, but
there does not appear to be any such indication in the data.

\section{Discussion and Conclusions}
We have presented a phenomenological model for the evolution of O abundance
relative to Fe. In this model, the first very massive stars formed from
Big Bang debris provided a prompt inventory of Fe 
([Fe/H]$_{\rm P}\sim -3$) and  O,
thus defining an initial state of the ISM. Subsequent
O enrichment was provided by SNII that are related to regular massive stars,
while subsequent Fe enrichment was provided by SNII followed by additional
contribution from SNIa. 
Prior to the onset of SNIa Fe contribution,
the evolution of (O/Fe) as a function of 1/(Fe/H) is along
a straight line segment (Fig. 1).
Subsequent evolution is governed by the rate
per unit mass of gas for Fe production 
by SNIa relative to SNII (Fig. 1). Based on the observations of
Israelian et al. (1998) and Boesgaard et al. (1999),
the value (O/Fe)$_{\rm P}$
resulting from the prompt production 
by the first very massive stars is $\sim 20$~(O/Fe)$_\odot$.
Comparison of our model with the same observations
indicate that SNIa began to contribute Fe
at [Fe/H]~$\sim -1$ and were responsible for $\sim 2/3$ of the solar Fe
inventory, consistent with the results of previous studies
(e.g., Timmes et al. 1995) and with the assumptions of Wasserburg \& Qian
(2000a,b). The average production of O relative to Fe for SNII is
found to be (O/Fe)$_{\rm II}\sim 3$~(O/Fe)$_\odot$, which is $\sim 7$
times lower than that for the first very massive stars.

The model 
developed here neither depends on nor gives numerical values of the rates
per unit mass of gas for O and Fe production by SNII, 
$d\,{\rm (O/H)}_{\rm II}/d\,t$ and $d\,{\rm (Fe/H)}_{\rm II}/d\,t$. 
Likewise, the model only involves $\gamma(t)$,
the rate per unit mass of gas for Fe production by SNIa
relative to SNII. Taking the Galaxy for example, we can obtain
the numerical values of $d\,{\rm (O/H)}_{\rm II}/d\,t$ and 
$d\,{\rm (Fe/H)}_{\rm II}/d\,t$ by 
introducing the time for solar system formation 
$t_{\rm SSF}\approx 10^{10}$~yr
(eqs. [\ref{odot}] and [\ref{rfe2}]). The absolute yields of O and Fe
from each SNII can also be obtained by further introducing the SNII rate 
per unit mass of gas and the total mass of gas at $t_{\rm SSF}$.
The progenitor stars of SNII have very short lifetimes so that
an equilibrium between the birth and death of these stars is 
established on a correspondingly short timescale.
If the rate for formation of these stars is proportional to
the mass of gas, then the SNII rate per unit mass of gas can be considered
as constant over Galactic history. This is the justification for
our earlier assumption
of a constant rate per unit mass of gas for Fe production by SNII 
(eq. [\ref{rfe2}]). At the present time, 
the Galactic SNII rate is $\sim (30\ {\rm yr})^{-1}$, 
corresponding to a total gas mass of $\sim 10^{10}\,M_\odot$.
Assuming that all SNII have the same O yield, we find that each SNII
must produce $\sim 0.3\,M_\odot$ of O in order to enrich the gas with
a solar O mass fraction ($\approx 9.6\times 10^{-3}$) at 
$t_{\rm SSF}\approx 10^{10}$~yr (eq. [\ref{xe}]).

In the model presented here, we have assigned the O and Fe abundances at
[Fe/H]~$\sim -3$ to the prompt production by the first very massive stars
formed from Big Bang Debris and have rejected the possibility that these 
could be produced by SNII. Indeed, although the value of (O/Fe)$_{\rm P}$ 
found here is $\sim 7$ times larger than the average relative production of 
O to Fe that we associate with SNII, it is still possible to attribute the
large (O/Fe) value at [Fe/H]~$\sim -3$ to contributions from a subset of
SNII that produce O but little Fe. The key argument that we 
have used in rejecting this possibility comes from the results on the
production of $r$-process elements and Fe by SNII that were obtained from 
meteoritic data and observations of metal-poor stars. We give a short 
summary of these results below as they will be used to place bounds on the 
O and Fe enrichment to be expected from the early occurrence of SNII in 
comparison with the prompt inventory (O/H)$_{\rm P}$ and (Fe/H)$_{\rm P}$.

Meteoritic data on the inventory of
radioactive $^{129}$I and $^{182}$Hf in the early solar system require
at least two distinct kinds of SNII as sources for the $r$-process
(Wasserburg, Busso, \& Gallino 1996; Qian, Vogel, \& Wasserburg 1998;
Qian \& Wasserburg 2000). These are the high-frequency SNIIH events
responsible for heavy $r$-process elements with mass numbers $A>130$
(e.g., Ba and Eu; ``H'' for ``high-frequency'' and ``heavy $r$-process 
elements'') and the low-frequency SNIIL events dominantly 
producing light $r$-process elements with $A\leq 130$ (e.g., Ag; ``L'' for
``low-frequency'' and ``light $r$-process elements''). 
The timescales required for replenishment of the appropriate radioactive 
nuclei in an average ISM by the SNIIH and SNIIL events are $\sim 10^7$~yr 
and $\sim 10^8$~yr, respectively. These timescales can be explained by
considering the mixing of the ejecta from an individual SNII with the ISM.
The amount of ISM to mix the SNII ejecta is the mass swept up by
a supernova remnant (SNR). The expansion of an SNR can be roughly
described by an energy-conserving phase followed by a momentum-conserving
one (e.g., Thornton et al. 1998). This expansion results in a total
swept-up mass of $\sim 2E_{\rm SN}/(v_{\rm tr}v_{\rm f})
\sim 3\times 10^4\,M_\odot$, where $E_{\rm SN}\sim 10^{51}$~erg is the
supernova explosion energy, $v_{\rm tr}\sim 200$~km~s$^{-1}$ is the
velocity at the transition between the two phases of expansion, and
$v_{\rm f}\sim 15$~km~s$^{-1}$ is the final velocity typical of the
random motion in the ISM. For Galactic rates of 
$\sim (30\ {\rm yr})^{-1}$ for H events and $\sim (300\ {\rm yr})^{-1}$
for L events in a total gas mass of
$\sim 10^{10}\,M_\odot$, we obtain that the rates of SNIIH and SNIIL 
events in an ISM of 
$\sim 3\times 10^4\,M_\odot$ are $\sim (10^7\ {\rm yr})^{-1}$
and $\sim (10^8\ {\rm yr})^{-1}$, respectively, as required by
replenishment of the appropriate radioactive nuclei in this ISM.

Observations of metal-poor stars in the Galactic halo
(McWilliam et al. 1995; McWilliam 1998; Sneden et al. 1996, 1998)
show that there is wide dispersion in the Ba abundance
$\log\epsilon({\rm Ba})\equiv\log({\rm Ba/H})+12$
at [Fe/H]~$\sim -3$ (region b in Fig. 3). 
This dispersion in the Ba abundance ($\sim 2$ dex) at a nearly constant 
[Fe/H] led Wasserburg \& Qian (2000a) 
to conclude that the SNIIH events cannot
produce a significant amount of Fe. The Ba abundance 
becomes extremely low for [Fe/H]~$<-3$ and there is no 
correlation with Fe in this region (region a in Fig. 3), indicating that there 
was some Fe production prior to the occurrence of SNII.
Observations (Gratton \& Sneden 1994; see also McWilliam 
et al. 1995) also show that
there is a correlation between the abundances of Ba and Fe at
[Fe/H]~$\gtrsim -2.5$ (region c in Fig. 3).  

Estimates of the Ba abundance resulting from a single SNIIH event
(Qian \& Wasserburg 2000; Wasserburg \& Qian 2000a) 
are shown as the band labeled ``1 SNIIH''
in Figure 3. The low Ba abundances in stars with
[Fe/H]~$\sim -4$ to $-3$ (region a in Fig. 3) compared with this band
and the sharp increase in the Ba abundance
at [Fe/H]~$\sim -3$ led Wasserburg \& Qian (2000a) to conclude
that a source other than SNII must exist to produce Fe (along with other
elements such as C, N, O, Mg, and Si) at [Fe/H]~$\lesssim -3$.
They attributed this source to the first very
massive stars (with masses of $\gtrsim 100\,M_\odot$) 
formed from Big Bang debris.
They further argued that formation of regular stars 
(with masses of $\sim 1$--$60\,M_\odot$)
could not occur until later when [Fe/H]~$\sim -3$ was reached.
In this later regime, regular stars with masses of 
$\gtrsim 10\,M_\odot$ explode as SNII.
The more frequent occurrence of SNIIH events with no Fe production over
a period of $\lesssim 10^8$~yr gives rise to 
the sharp increase in the Ba abundance at [Fe/H]~$\sim -3$. This is 
followed by the occurrence of low-frequency Fe-producing SNIIL events.
Contribution to Fe from SNIa only takes place at 
[Fe/H]~$\gtrsim -1$. Enrichment in Fe of metal-poor stars with
$-1\gtrsim$~[Fe/H]~$\gtrsim -2.5$ must be provided by the SNIIL events.
For a Galactic rate of $\sim (300\ {\rm yr})^{-1}$
corresponding to a total gas mass of $\sim 10^{10}\,M_\odot$,
each SNIIL must produce $\sim 0.1\,M_\odot$ of Fe in order to
enrich the gas with $\sim 1/3$ of a solar Fe mass fraction 
($\approx 1.2\times 10^{-3}$)
at $t_{\rm SSF}\approx 10^{10}$~yr (eq. [\ref{xe}]). 
Diluting this amount of Fe in an ISM of 
$\sim 3\times 10^4\,M_\odot$ that mixes with the ejecta of a single 
SNII event results in
[Fe/H]~$\sim -2.5$. Thus the correlation 
between the abundances of Ba and Fe at [Fe/H]~$\gtrsim -2.5$ shows that
the variation of Fe yields between SNIIH and SNIIL events 
is smoothed out over a timescale of $\gtrsim 10^8$~yr.

While the SNIIH events would produce oxygen, 
they will not be sufficient to provide the large O inventory 
at [Fe/H]~$\sim -3$. 
Diluting the O yield of $\sim 0.3\,M_\odot$ from an SNIIH
event into an ISM of $\sim 3\times 10^4\,M_\odot$ would result in
(O/H)~$\sim 10^{-3}$~(O/H)$_\odot$ in this mixture.
We note that two stars with Ba abundances close to our 
estimate for a single SNIIH event and with essentially the 
same values of [Fe/H]~$\sim -3$ are both observed to have
(O/H)~$\approx 10^{-2}$~(O/H)$_\odot$, which is a factor of $\sim 10$ 
greater than our estimate for a single SNIIH event
(HD122563 has $\log\epsilon({\rm Ba})=-1.54$ and [Fe/H]~$=-2.74$, 
and BD $-18^\circ$5550 has $\log\epsilon({\rm Ba})=-1.72$ and 
[Fe/H]~$=-2.91$; Westin et al. 2000; McWilliam 
1998; Cavallo, Pilachowski, \& Rebolo 1997).

We also note that the prompt value (O/Fe)$_{\rm P}\sim 20$~(O/Fe)$_\odot$ 
at [Fe/H]~$\sim -3$ assigned by us is based on the observations of
BD +23$^\circ$3130 with [Fe/H]~$=-2.9$ (Israelian et al. 1998), 
BD +03$^\circ$740 with [Fe/H]~$=-3$, 
and BD $-13^\circ$3442 with [Fe/H]~$=-3.13$ 
(Boesgaard et al. 1999; Israelian et al. 2000) but there are 
no relevant Ba data. This value of (O/Fe)$_{\rm P}$ corresponds to
(O/H)$_{\rm P}\sim 2\times 10^{-2}$~(O/H)$_\odot$, which could be
explained by the O contribution from $\sim 20$ SNIIH events at
$\sim 0.3\,M_\odot$ of O per event.
If this were the case, then the relevant stars should have 
high $\log\epsilon({\rm Ba})$ values. 
Future measurements of Ba abundances in these
stars are needed to identify possible SNIIH contribution to
the observed O abundances and to determine the prompt O inventory 
more accurately. However, from the above arguments based on the
observations of HD 122563 and BD $-18^\circ$5550, we conclude that 
the origin of the O inventory at [Fe/H]~$\lesssim -3$ must be assigned 
to the prompt source.

Based on the assignment of 
(O/Fe)$_{\rm P}\sim 20$~(O/Fe)$_\odot$, the average production ratio
of O to Fe for the first very massive stars is $\sim 560$ by number
[(O/Fe)$_\odot\approx 28$]. As the amount
of O production is much less than the mass of the star,
this means that the typical Fe yield from a zero-metallicity
star of mass $M$ must be $\ll 0.6\,M_\odot\times(M/100\,M_\odot)$. 
In considering the prompt enrichment of the ISM
by the very massive stars, it is necessary 
to take several effects into account. 
The universe at the epoch for formation
of these stars was probably dominated by gas. We do not know 
how many very massive stars contributed to the prompt inventory.
Explosions of these stars are likely
to be much more energetic than supernova explosions. So the 
mass of ISM for diluting the ejecta from the explosion of 
a very massive star is expected to be $\gg 3\times 10^4\,M_\odot$.
With an O yield of $\sim 0.1\,M$ for a very massive star
of mass $M$ and with the number $N_{\rm VMS}$ of such stars 
contributing to the prompt
O inventory of (O/H)$_{\rm P}\sim 2\times 10^{-2}$~(O/H)$_\odot$, 
this dilution mass can be estimated as 
$\sim 5\times 10^4\,M_\odot(M/100\,M_\odot)\times N_{\rm VMS}$.
The expectation that this mass should be $\gg 3\times 10^4\,M_\odot$
leads us to conclude that the prompt inventory must have been provided
by $N_{\rm VMS}\gg 1$ very massive stars. That is, many very massive
stars are required to provide (O/H)$_{\rm P}$.
Clearly, further studies are needed to understand the formation,
evolution, nucleosynthesis, and explosion of these stars.

To determine the relationship between (O/Fe) and (Fe/H) in the Galaxy, 
Israelian et al. (1998) and Boesgaard et al. (1999) used the OH lines of 
metal-poor stars with a wide range of (Fe/H). 
Their results represent the most complete 
and self-consistent data set available. There is some disagreement 
between different workers on the evolution of O abundance relative to Fe,
in part due to the choice of spectral lines used to determine the O
abundance (e.g., Barbuy 1988). While the general model developed 
here must stand on its own, the numerical value of (O/Fe)$_{\rm P}$ for the
prompt inventory is dependent on the observational results of
Israelian et al. (1998) and Boesgaard et al. (1999).

Up to this point, we have assumed that the ISM consists of gas only
and that the abundances in an individual star are representative of 
the average gas at the birth of the star. In fact, the Galactic ISM 
consists of both gas and dust, with dust making up $\sim 1\%$ of 
the total mass at present. Certain 
elements such as Fe in the ISM are greatly depleted in the gas phase and
are considered to be condensed into dust.
So long as stars sample the composition of the bulk ISM 
including both gas and dust, we can still interpret stellar observations 
in terms of Galactic chemical evolution as done here because
dust is converted into gas in stars.
In contrast, when applying the model developed here to 
damped Ly$\alpha$ systems, it must be further 
assumed that Fe and O are not substantially fractionated relative 
to each other in the protogalactic ISM of such systems. 
If for example, Fe were dominantly in dust 
and O in gas in the ISM of damped Ly$\alpha$ systems
(as is the case in many galaxies of about the present epoch), then a 
straightforward application of the model to such systems would give 
completely misleading interpretation of the data, which only sample
elements in the gas phase. 

The most accurate measurement of O abundance in a damped Ly$\alpha$ 
system so far was reported by Molaro et al. (2000). They gave 
[O/H]~$=-1.85\pm 0.1$ corresponding to [O/Fe]~$=0.19\pm 0.14$
for a damped Ly$\alpha$ system with [Fe/H]~$=-2.04\pm 0.1$ at a redshift
$z=3.39$. These data
do not fit the trend exhibited by the
Galactic data shown in Figure 2d. The data on other elements (e.g., Cr and Zn)
in the above damped Ly$\alpha$ system also suggest that there is 
no effect of dust depletion (Molaro et al. 2000). We note that 
the observed value of [O/H] in this system is close to the prompt O
inventory assigned here based on the Galactic data while the value of
[Fe/H] is $\sim 10$ times higher than the prompt Fe inventory.
We cannot offer a simple explanation for these O and Fe abundances.

Substantial efforts have been made to establish the dust-to-gas ratio
for damped Ly$\alpha$ systems at redshifts $z\gtrsim 1$. 
Estimates have been made for the possible dust depletion
of condensible elements (e.g., Si, Cr, and Fe) relative to more volatile 
ones (e.g., Zn) that are considered to remain in the gas phase. 
Pettini et al. (1997) have argued that there is some 
significant dust depletion of condensible elements
(by a factor of $\sim 2$) in the
damped Ly$\alpha$ systems at very early epochs. However, there is strong 
disagreement between various workers (e.g., Lu et al. 1996;
Prochaska \& Wolfe 2000) regarding the degree of dust depletion 
in these systems.
Wasserburg \& Qian (2000b) assumed that dust depletion is not 
a dominant effect on the chemical evolution of damped Ly$\alpha$ systems
at $z\gtrsim 1$. There is clear evidence for large amounts of 
dust in a variety of galaxies at high $z$. We cannot 
reconcile the assumption of low dust-to-gas ratios for damped 
Ly$\alpha$ systems with these observations. The cause of the 
transition to the high dust content for galaxies of about the present 
epoch is not understood unless the later dust contribution of 
Asymptotic Giant Branch (AGB) stars plays a major role.

It is expected that the first very massive stars formed from Big Bang 
debris produce C/O ratios of $\ll 1$ (Heger et al. 2000). 
The dominance of O over C should cause all the C to 
be converted into CO, leaving an excess of O in the ISM. 
This would prevent the formation of C grains and also the 
formation of polyaromatic hydrocarbon (PAH) 
as well as other hydrocarbon molecules. A 
strict upper limit to the dust content of the ISM at 
[Fe/H]~$\sim -3$ can be estimated by taking the total abundance
of the major condensible elements Mg, Si, and Fe to be 
$\sim 10$ times the Fe abundance (see below). As no C dust should 
be formed, dust would constitute at most $\sim 10^{-5}$
of the total mass of the ISM at [Fe/H]~$\sim -3$. For a typical
dust size $d$, the obscuration due to this presence of dust is 
at most at the level of $\sim 10^{-3}
(0.1\ \mu{\rm m}/d)[N({\rm H\ I})/10^{21}\ {\rm cm}^{-2}]$
in a system of neutral H column density $N({\rm H\ I})$.

By attributing the O inventory at [Fe/H]~$\sim -3$ to the prompt production
by the first very massive stars formed from Big Bang debris, the model
presented here also predicts that the Mg and Si abundances in stars with
(O/Fe)~$\sim$~(O/Fe)$_{\rm P}$ and [Fe/H]~$\sim -3$ should to a large
extent reflect the prompt inventory of Mg and Si. Based on the observations
of McWilliam et al. (1995), we obtain 
(Mg/H)$_{\rm P}\sim 3\times 10^{-3}$~(Mg/H)$_\odot$ and
(Si/H)$_{\rm P}\sim 10^{-2}$~(Si/H)$_\odot$, corresponding to
(Mg/Fe)~$\sim 3$~(Mg/Fe)$_\odot$ and (Si/Fe)~$\sim 10$~(Si/Fe)$_\odot$
for stars with [Fe/H]~$\sim -3$. We also consider that
the abundances in the prompt inventory represent the critical condition
for transition from formation of very massive stars to regular ones.
Taking into account the radiative properties of H, O, Mg, Si, and Fe,
we predict that the abundances (O/H)~$\sim 2\times 10^{-2}$~(O/H)$_\odot$, 
(Mg/H)~$\sim 3\times 10^{-3}$~(Mg/H)$_\odot$, 
(Si/H)~$\sim 10^{-2}$~(Si/H)$_\odot$, and
(Fe/H)~$\sim 10^{-3}$~(Fe/H)$_\odot$ in the prompt inventory should give
sufficient ``metals'' to permit cooling and fragmentation of large 
condensing gas clouds to form stars with masses of $\sim 1$--$60\,M_\odot$.
 
\acknowledgments
We would like to thank Garik Israelian for supplying the data of
Israelian et al. (1998) and Boesgaard et al. (1999) on the same
scale of stellar parameters. Comments by the referee, F. X. Timmes, 
were very helpful in improving the presentation.
This work was supported in part by the Department of Energy under grant
DE-FG02-87ER40328 to Y.-Z. Q. and by NASA under grant
NAG5-4083 to G. J. W., Caltech Division Contribution No. 8732(1063).

\clearpage
\figcaption{Illustration of the evolution of (O/Fe) as a function of
(Fe/H)$_{\rm P}$/(Fe/H). Here (O/Fe)$_{\rm P}$ 
represents the production ratio for the prompt source, (O/Fe)$_{\rm II}$
for SNII, and (O/Fe)$_{\rm I}=0$ for SNIa. The initial state P results
from the production by the prompt source. State A corresponds to
the onset of SNIa Fe contribution. The general evolution is
described by a three-component mixing model involving the initial
state P, SNII, and SNIa. Prior to reaching state A, the evolution
is along the line segment PII, with point II representing
the state to be reached after an infinite number of SNII. 
(a) If SNII contributions were to cease at state A, 
subsequent evolution would be along the line segment PI, 
with point I representing the state to be reached after an infinite 
number of SNIa. On the other hand, 
if the rate per unit mass of gas for Fe production by
SNIa relative to SNII were a constant $\gamma$, the evolution 
beyond state A would be along the line segment AC, with point C
representing the state to be reached after an infinite number of SNII 
and SNIa. Note that the line segments AII and AI correspond to the
limits $\gamma=0$ and $\infty$. (b) In general, the rate per unit mass 
of gas for Fe production by SNIa relative to SNII is a function of time
$\gamma(t)$. So the evolution beyond state A follows a
curve between the line segments AII and AI. The tangent at an arbitrary
point B on the curve is the same as the slope of the line segment BC$'$,
with point C$'$ representing the state to be reached after an
infinite number of SNII and SNIa for a constant $\gamma$
equal to the value $\gamma_{\rm B}$ at point B. Thus if $\gamma(t)$
monotonically increases with time, the curve drops more
and more steeply in approaching (Fe/H)$_{\rm P}$/(Fe/H)~$=0$. 
(c) In general,
the curve is concave downward if $\gamma(t)$ increases
with time ($\dot\gamma>0$) and is concave upward if $\gamma(t)$ decreases
with time ($\dot\gamma<0$). In any case, the value of (O/Fe) to be
reached after an infinite number of SNII and SNIa is determined by
the asymptotic value $\gamma_\infty$ of $\gamma(t)$.}

\figcaption{Comparison of the model for the evolution of O abundance
relative to Fe with the data for metal-poor stars in the
Galaxy (filled circles: Israelian et al. 1998; squares: Boesgaard et al.
1999). Note the change in the trend of the data at 
(Fe/H)$_{\rm P}$(Fe/H)$\sim 0.02$ highlighted in the inset of (a).
The data for
Galactic disk stars (crosses: Edvardsson et al. 1993)
are included in (c) and (d) to indicate the trend near and
beyond the solar point (circle). 
The solid curve is obtained from the model by taking
(O/Fe)$_{\rm P}=21$~(O/Fe)$_\odot$,
(Fe/H)$_{\rm P}=10^{-3}$~(Fe/H)$_\odot$,
(Fe/H)$_{\rm A}=0.05$~(Fe/H)$_\odot$
[(Fe/H)$_{\rm P}$/(Fe/H)$_{\rm A}=0.02$],
$\alpha=1/3$, and $\hat\tau/t_{\rm SSF}=0.1$.
The long-dashed curve in (c) is
obtained by taking $\hat\tau/t_{\rm SSF}=1$ and the short-dashed curve
by taking (Fe/H)$_{\rm A}=0.1$~(Fe/H)$_\odot$ 
[(Fe/H)$_{\rm P}$/(Fe/H)$_{\rm A}=0.01$] while keeping all the other
parameters the same as for the solid curve.}

\figcaption{Data (asterisks: Gratton \& Sneden 1994; squares: 
McWilliam et al. 1995; McWilliam 1998; triangles: Sneden et al. 1996, 1998)
on $\log\epsilon({\rm Ba})$ versus [Fe/H] for metal-poor
stars in the Galaxy. The band labeled ``1 SNIIH'' shows estimates of the 
$\log\epsilon({\rm Ba})$ value resulting from a single SNIIH event.
The wide dispersion in $\log\epsilon({\rm Ba})$ 
at [Fe/H]~$\sim -3$ (region b) indicates that the SNIIH events
cannot produce a significant amount of Fe. The correlation between the 
abundances of Ba and Fe at [Fe/H]~$\gtrsim -2.5$ (region c) results
from the mixture of the SNIIH and the Fe-producing SNIIL events.
The low Ba abundances in stars with $-4\lesssim$~[Fe/H]~$\lesssim -3$
(region a) compared with the contribution from a single SNIIH event 
and the sharp increase of $\log\epsilon({\rm Ba})$ at [Fe/H]~$\sim -3$
(region b) suggest that a prompt source other than SNII must exist to 
produce Fe (along with other elements such as C, N, O, Mg, and Si) 
at [Fe/H]~$\lesssim -3$.}

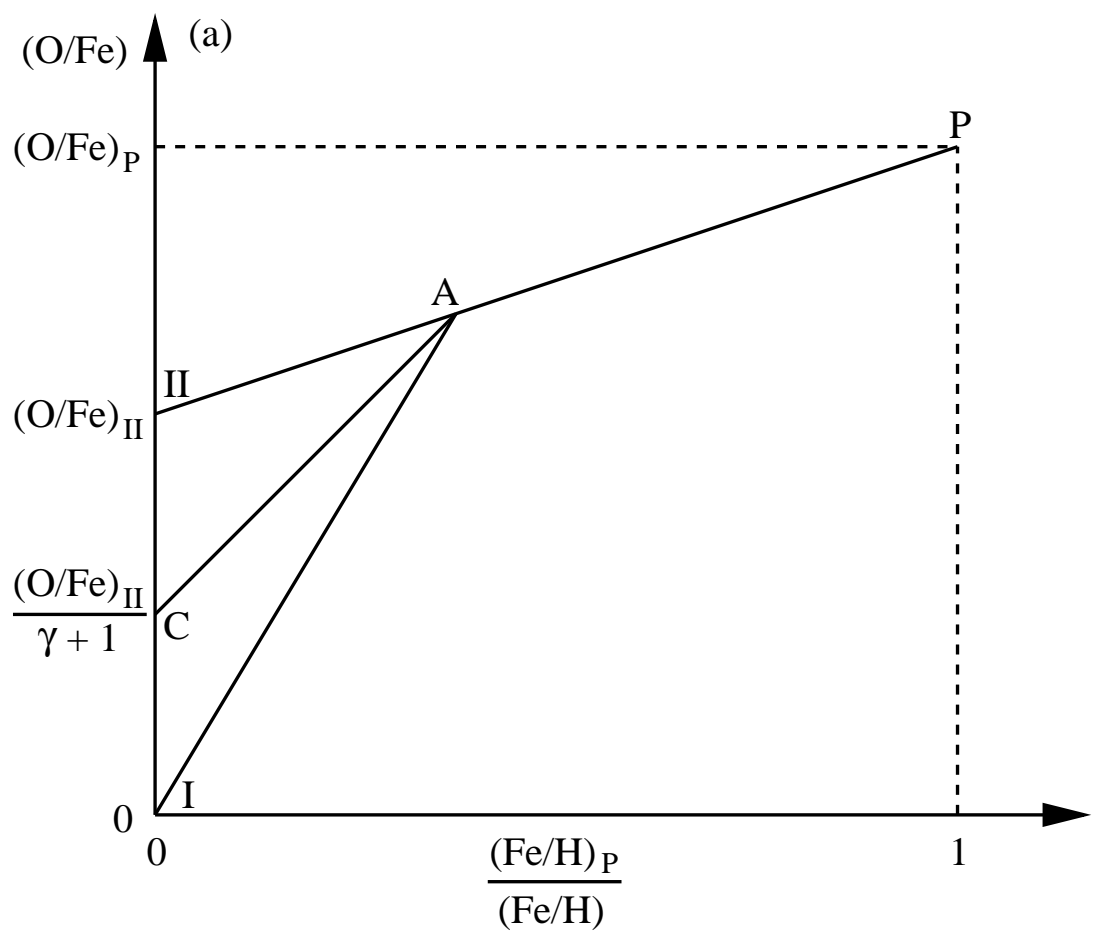
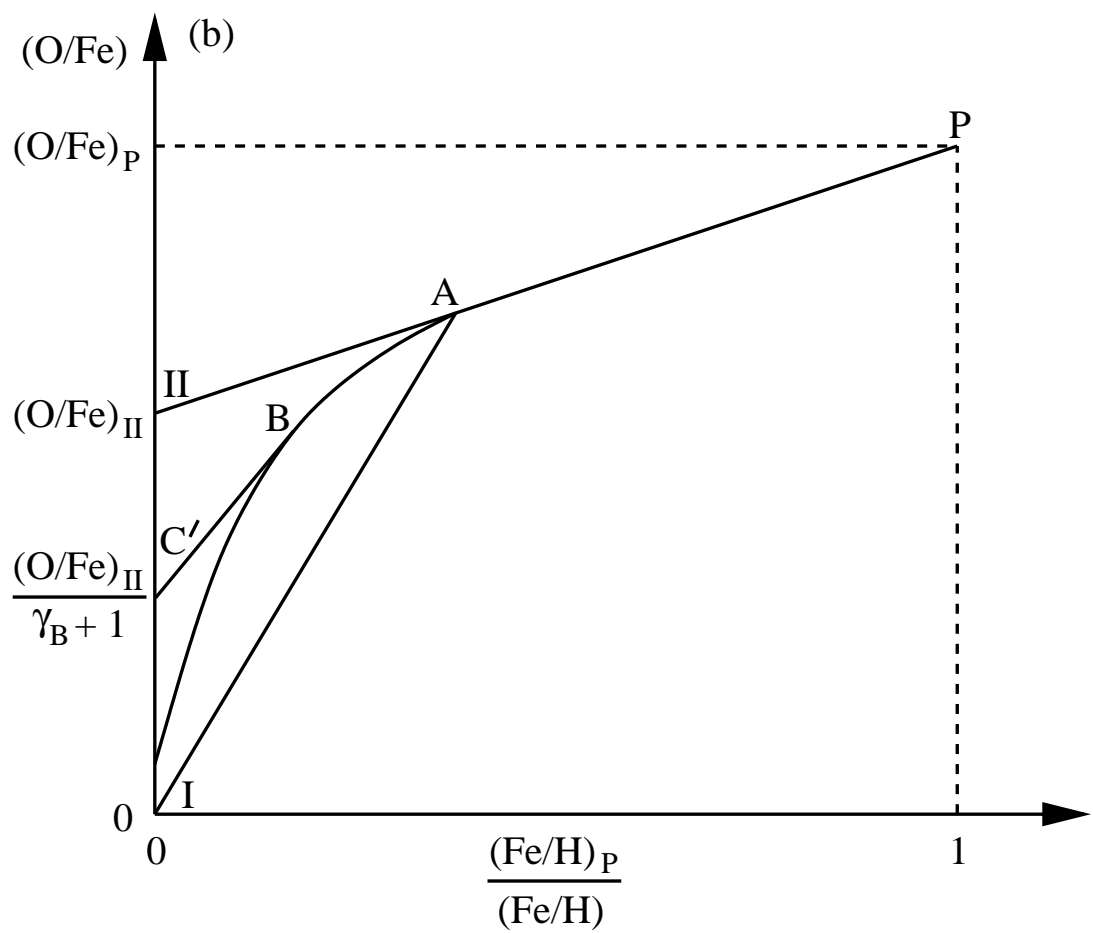
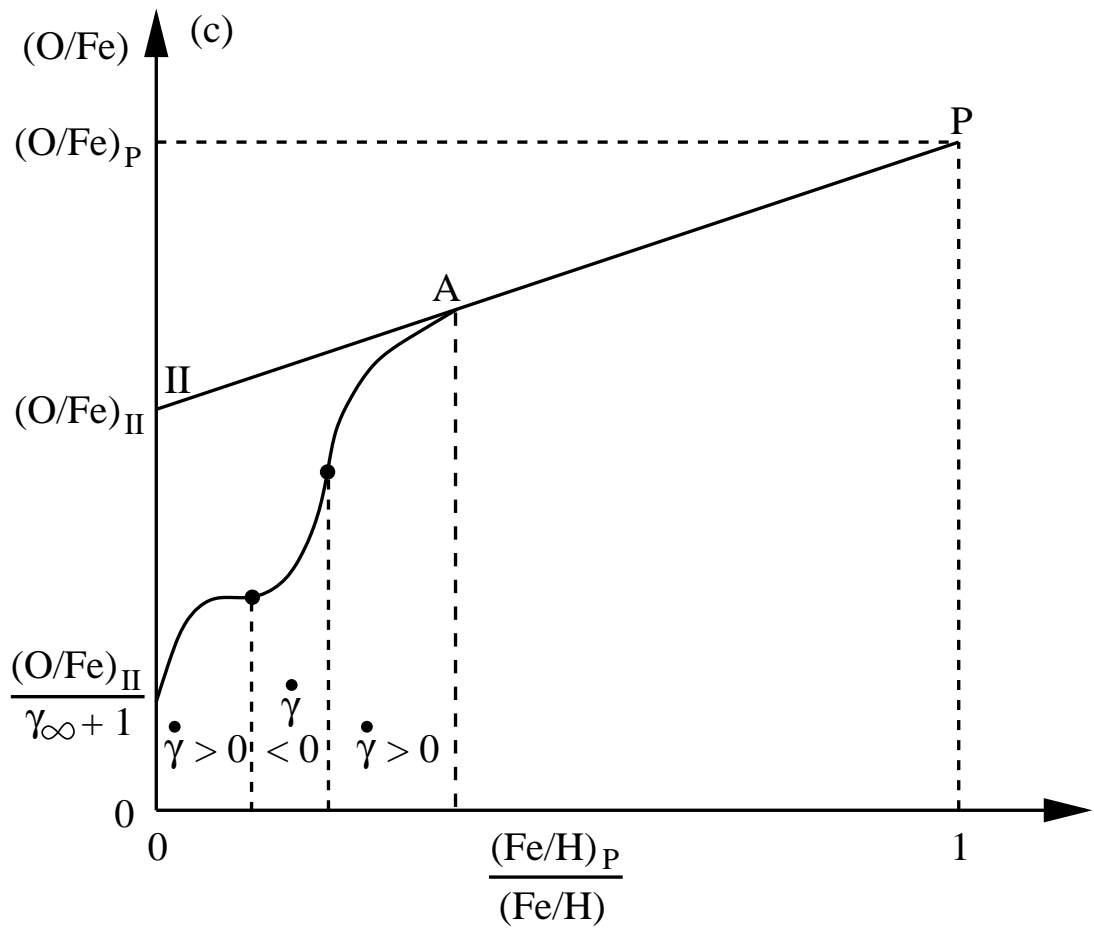
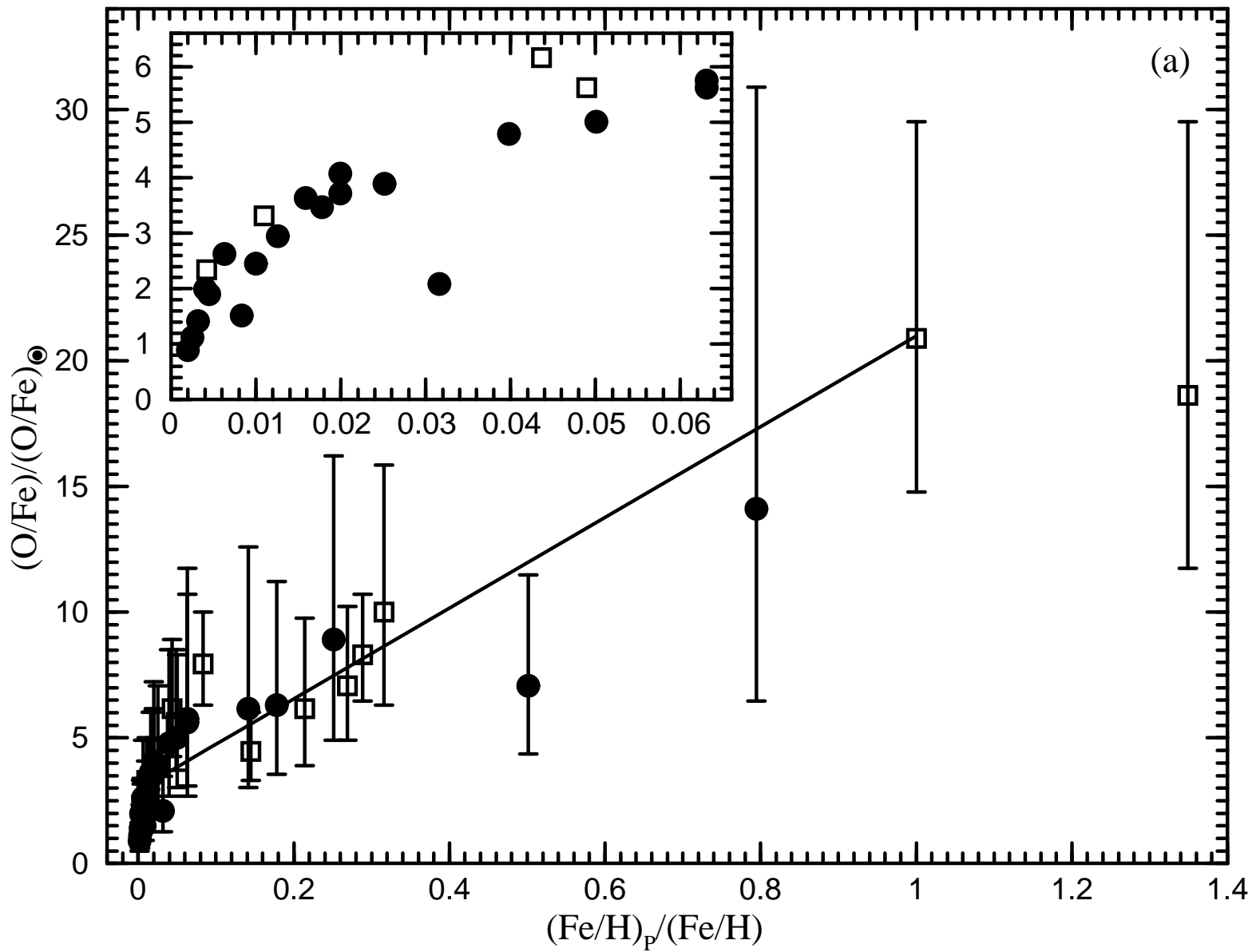
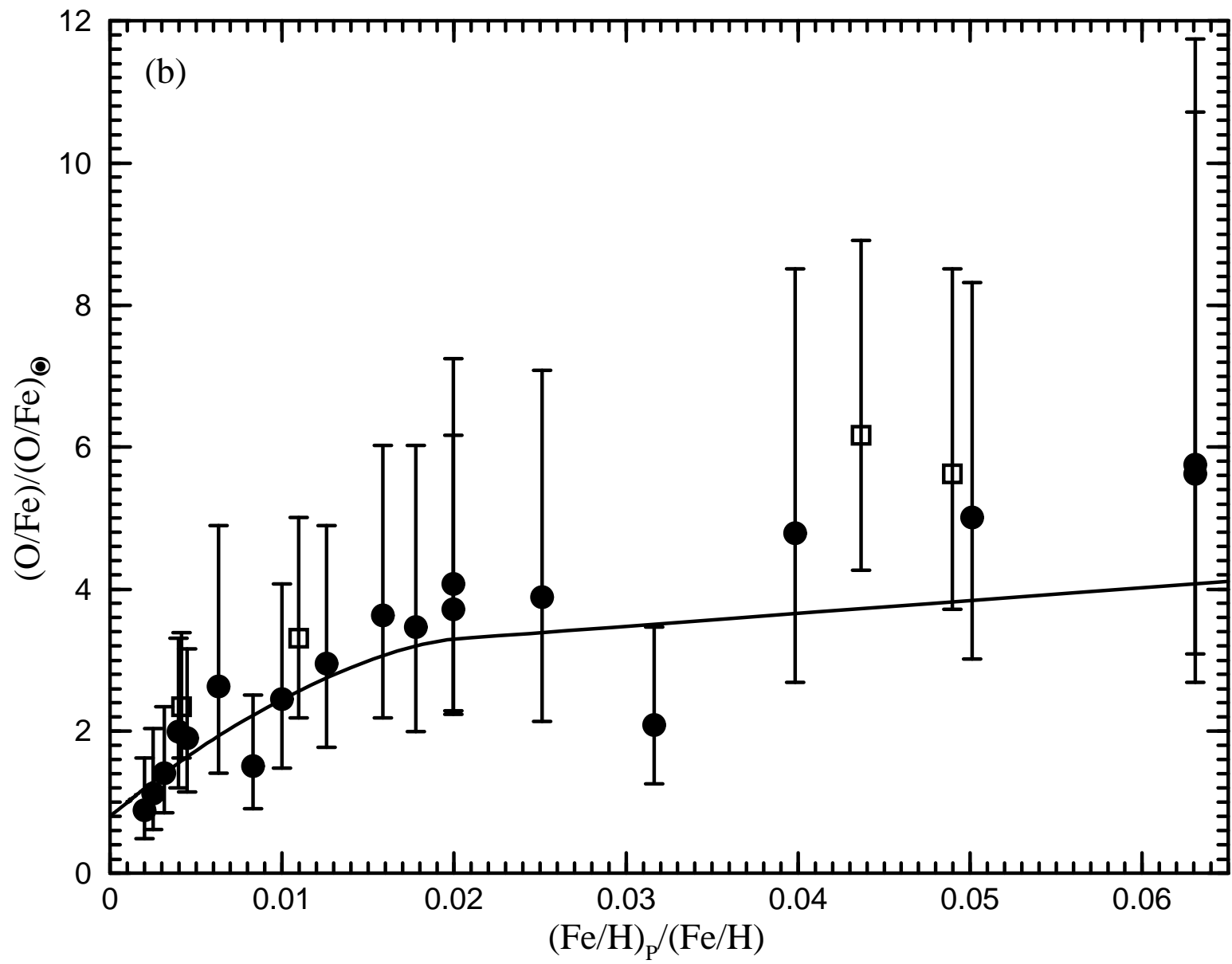
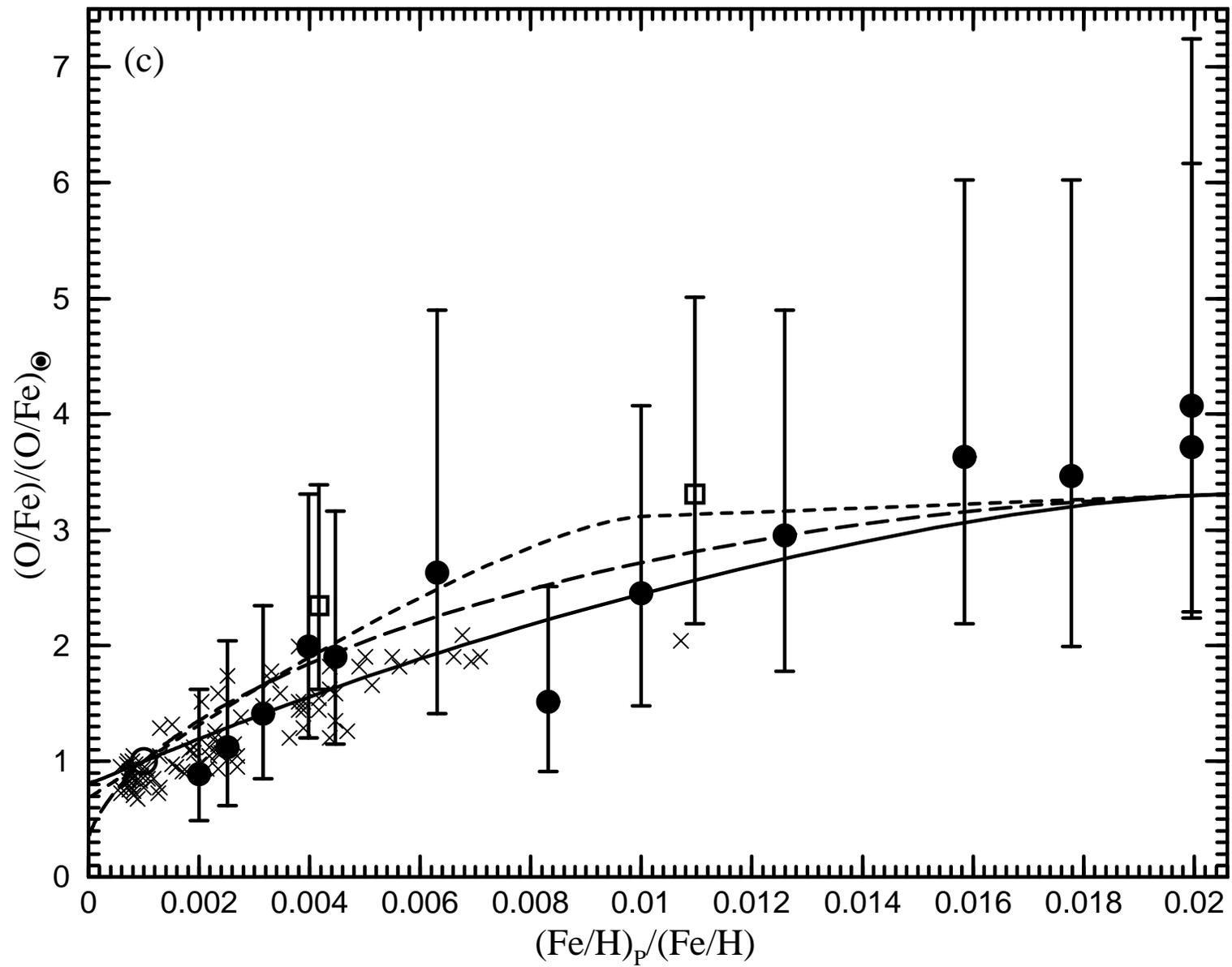
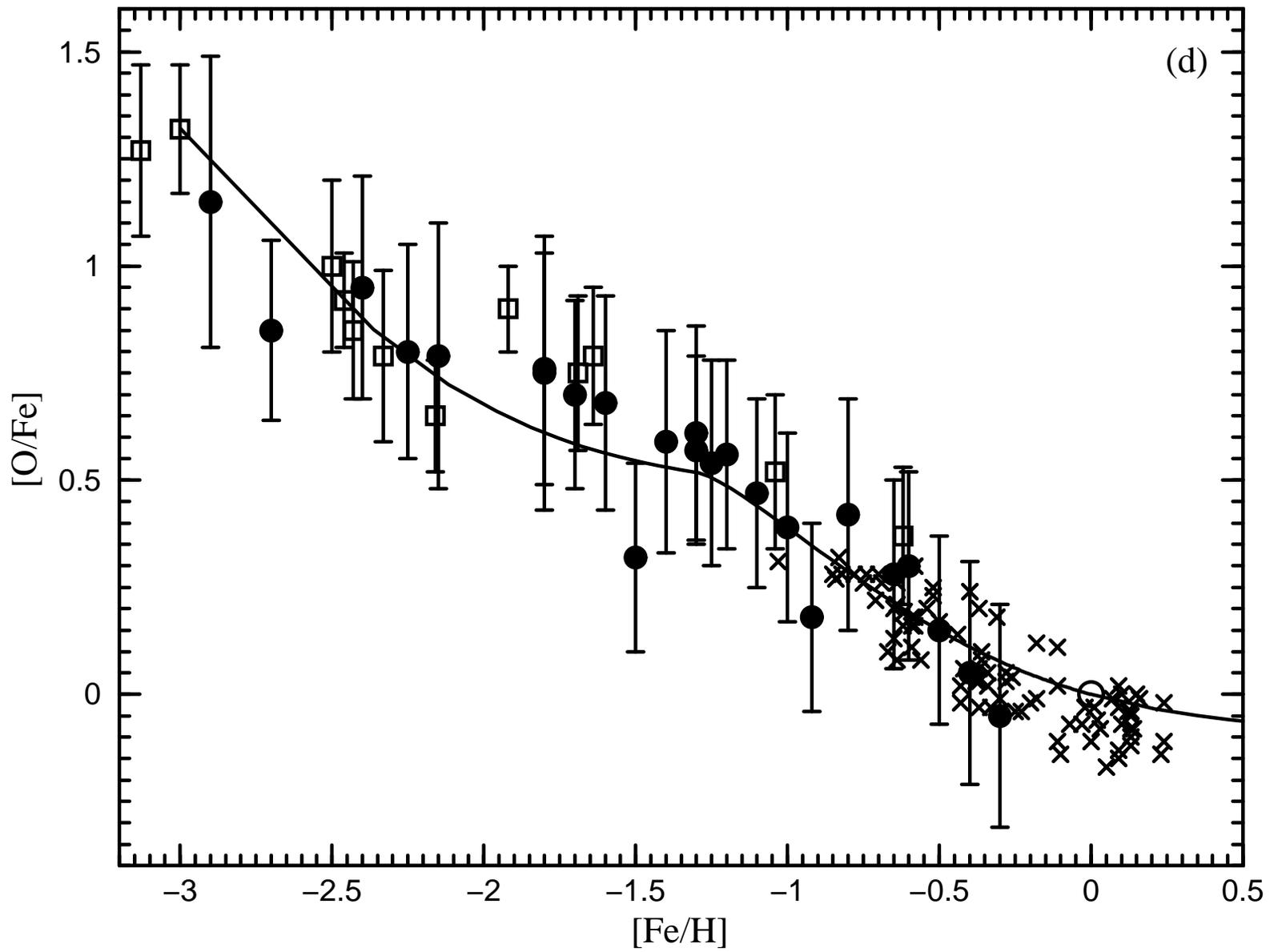
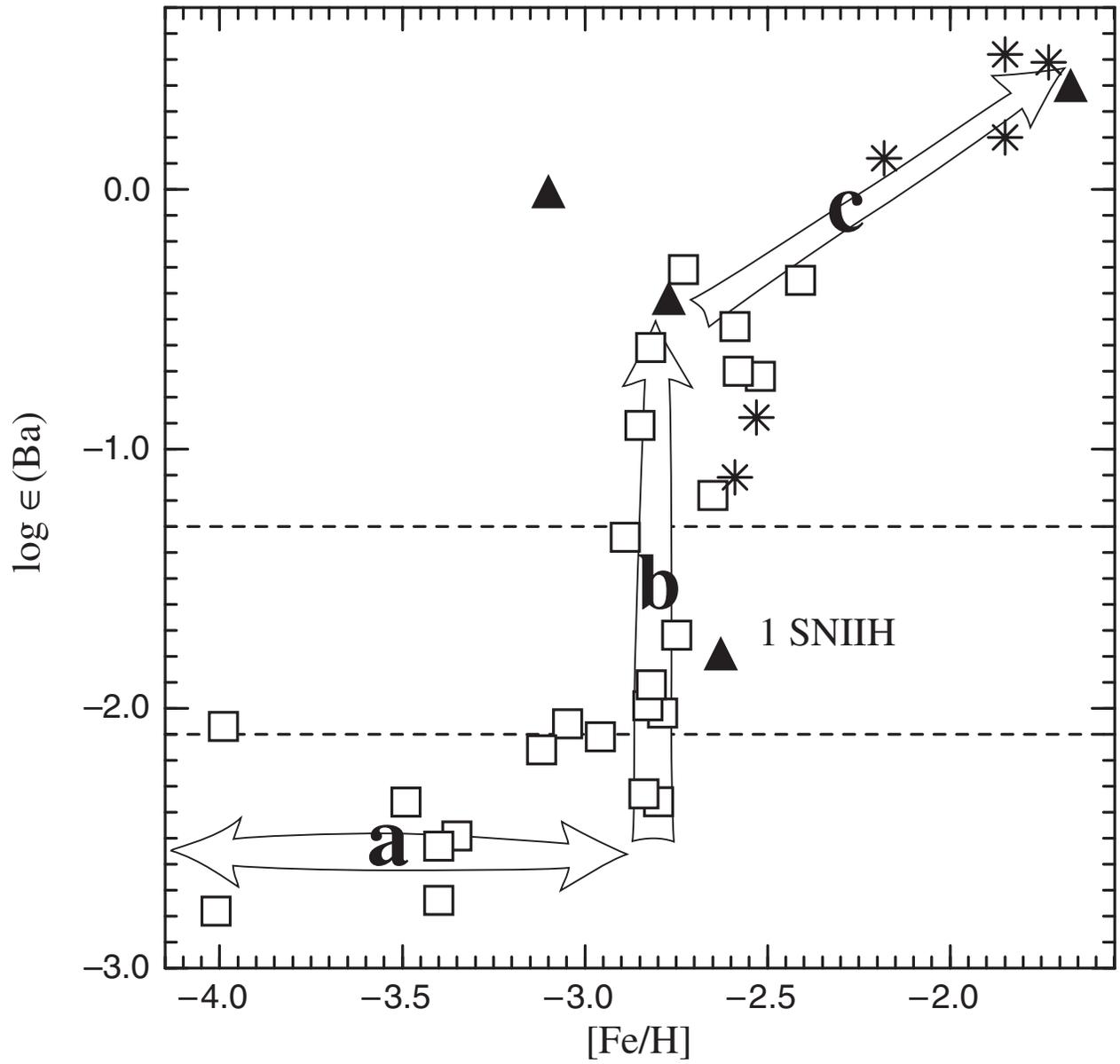
\end{document}